\documentclass[aps,prl,preprint,showpacs,showkeys,superscriptaddress,a4paper]{revtex4}

\newcommand{\degr}{$^{\circ}$}

\newcommand{\bm}[1]{\mbox{\boldmath $#1$}}
\newcommand{\fsi}{\emph{FSI}}
\usepackage{graphicx,amssymb,amsmath,color}

\usepackage{dcolumn}
\usepackage{bm}

\begin{document}

\title{Study of the $\bm{p+^6}$Li$\bm{\to\eta+^7}$Be reaction 11.3 MeV above threshold}

\author{A. Budzanowski}
\affiliation{Institute of Nuclear Physics, PAN, Krakow, Poland}

\author{A. Chatterjee}
\affiliation{Nuclear Physics Division, BARC, Bombay, India}

\author{P. Hawranek}
\affiliation{Institute of Physics, Jagellonian University, Krakow,
Poland}

\author{R. Jahn}
\affiliation{Helmholtz-Institut f\"{u}r Strahlen- und Kernphysik der
Universit\"{a}t Bonn, 53115 Bonn, Germany}

\author{V. Jha}
\affiliation{Nuclear Physics Division, BARC, Bombay, India}

\author{K. Kilian}
\affiliation{Institut f\"{u}r Kernphysik, Forschungszentrum J\"{u}lich,
J\"{u}lich, Germany}

\author{S. Kliczewski}
\affiliation{Institute of Nuclear Physics, PAN, Krakow, Poland}

\author{Da. Kirillov}
\affiliation{Institut f\"{u}r Kernphysik, Forschungszentrum J\"{u}lich,
J\"{u}lich, Germany}
\affiliation{Fachbereich Physik, Universit\"{a}t
Duisburg-Essen, Duisburg, Germany}

\author{Di. Kirillov}
\affiliation{Laboratory for High Energies, JINR Dubna, Russia}

\author{D. Kolev}
\affiliation{Physics Faculty, University of Sofia, Sofia, Bulgaria}

\author{M. Kravcikova}
\affiliation{Technical University, Kosice, Kosice, Slovakia}

\author{M. Lesiak}
\affiliation{Institut f\"{u}r Kernphysik, Forschungszentrum J\"{u}lich,
J\"{u}lich, Germany}
\affiliation{Institute of Physics, Jagellonian
University, Krakow, Poland}

\author{J. Lieb}
\affiliation{Physics Department, George Mason University, Fairfax,
Virginia, USA}

\author{H. Machner}
\email{h.machner@fz-juelich.de} \affiliation{Institut f\"{u}r
Kernphysik, Forschungszentrum J\"{u}lich, J\"{u}lich, Germany}
\affiliation{Fachbereich Physik, Universit\"{a}t Duisburg-Essen,
Duisburg, Germany}

\author{A. Magiera}
\affiliation{Institute of Physics, Jagellonian University, Krakow,
Poland}

\author{R. Maier}
\affiliation{Institut f\"{u}r Kernphysik, Forschungszentrum J\"{u}lich,
J\"{u}lich, Germany}

\author{G. Martinska}
\affiliation{P. J. Safarik University, Kosice, Slovakia}

\author{N. Piskunov}
\affiliation{Laboratory for High Energies, JINR Dubna, Russia}

\author{D. Proti\'c}
\affiliation{Institut f\"{u}r Kernphysik, Forschungszentrum J\"{u}lich,
J\"{u}lich, Germany}

\author{J. Ritman}
\affiliation{Institut f\"{u}r Kernphysik, Forschungszentrum J\"{u}lich,
J\"{u}lich, Germany}

\author{P. von Rossen}
\affiliation{Institut f\"{u}r Kernphysik, Forschungszentrum J\"{u}lich,
J\"{u}lich, Germany}

\author{B. J. Roy}
\affiliation{Nuclear Physics Division, BARC, Bombay, India}

\author{I. Sitnik}
\affiliation{Laboratory for High Energies, JINR Dubna, Russia}

\author{R. Siudak}
\affiliation{Institute of Nuclear Physics, PAN, Krakow, Poland}

\author{R. Tsenov}
\affiliation{Physics Faculty, University of Sofia, Sofia, Bulgaria}

\author{J. Urban}
\affiliation{P. J. Safarik University, Kosice, Slovakia}

\author{G. Vankova}
\affiliation{Institut f\"{u}r Kernphysik, Forschungszentrum J\"{u}lich,
J\"{u}lich, Germany}
\affiliation{Physics Faculty, University of Sofia,
Sofia, Bulgaria}

\collaboration{The COSY-GEM Collaboration}\noaffiliation





\date{\today}

\begin{abstract}
The cross section for the reaction $p+^6\text{Li}\to\eta+^7\text{Be}$ was measured at an excess energy of 11.28 MeV above threshold by detecting the recoiling $^7$Be nuclei. A dedicated set of focal plane detectors  was built for the magnetic spectrograph Big Karl and was
used for identification and four momentum measurement of the $^7$Be. A differential cross section of $\frac{d\sigma}{d\Omega}=(0.69\pm{0.20}\text{( stat.)}\pm 0.20\text{( syst.)})\text{ nb/sr}$ for the ground state plus the 1/2$^-$ was measured. The result is compared to model calculations.
\end{abstract}


%

\maketitle
\setcounter{equation}{0}

\section{Introduction}\label{sec:Introduction}
The possibility of the $\eta$ meson forming a quasi-bound state in a
nucleus was first raised by Haider and Liu~\cite{Haider_Liu86}. Such
a state could arise as a consequence of the strongly attractive
$\eta$-nucleon interaction that is driven by the $N^{*}(1535) S_{11}$
resonance. By using the $s$-wave $\eta N$ scattering length $a_{\eta
N}\approx (0.28 + 0.19i)$~fm, Bhalerao and Liu \cite{Bhalerao85}
found that the $\eta$ meson could form a
quasi-bound state with nuclei of mass number $A \ge
10$~\cite{Haider_Liu86}. Other groups found similar results when
starting from this relatively small value of $a_{\eta
N}$~\cite{Hayano99, Garcia-Recio02}. However,  Rakityansky et al. \cite{Rakityansky96} claimed that an $\eta$-nucleus quasi-bound state may exist for $A\geq2$, but widths of such
quasi-bound states could be small only for the $\eta^4$He system. Binding of the $\eta^4$He system was also found in \cite{Wycech95} and \cite{Scoccola98}. All calculations spanning a larger mass scale found that binding increases with increasing mass number.

In a recent study we found strong evidence that such a quasi bound state exists for $\eta\oplus^{25}$Mg by making use of a two nucleon transfer reaction $p+^{27}$Al$\to ^3$He+X at recoil free conditions \cite{Budzanowski09a}, i.e. the $^3$He carries the beam momentum and the bound $\eta$ is almost at rest. Then a second step occurs inside the nucleus $\eta+n\to \pi^-+p$ with the two charged particles being emitted almost back to back. These two charged particles were recorded with a dedicated large acceptance detector \cite{Betigeri07}.

Another approach to search for such quasi-bound $\eta$ nuclei is to study the final state interaction (\fsi) in two body final state reactions. Recently, two different experiments at COSY J\"{u}lich measured $\eta$ production in $pd\to {\eta\rm{^3He}}$ reactions very close to threshold with extremely high precision data \cite{Smyrski07, Mersmann07}. Whereas Smyrski et al. \cite{Smyrski07} claimed that only a scattering length is sufficient to describe the data, Mersmann et al. \cite{Mersmann07} found a better description when an effective range is also taken into account.
The result of the first group is
$a_{{}\rm{^3He}\eta }  = \left[ { \pm \left( {2.9 \pm 2.7} \right) + i \cdot \left( {3.2 \pm 1.8} \right)} \right]{\rm{fm}}$, while the second group reported
$a_{{}\rm{^3He}\eta }  = \left[ { \pm \left( {10.7 \pm 0.8_{ - 0.5}^{ + 0.1} } \right) + i \cdot \left( {1.5 \pm 2.6_{ - 0.9}^{ + 1.0} } \right)} \right]{\rm{fm}}
$
and
$
r_0  = \left[ {\left( {1.9 \pm 0.1} \right) + i \cdot \left( {2.1 \pm 0.2_{ - 0.0}^{ + 0.2} } \right)} \right]{\rm{fm}}$
for the effective range. However, Smyrski et al. did not include smearing due to the experimental resolution in the calculation. The nearby pole hypothesis is confirmed by a careful study of the energy dependence of the angular variation \cite{Wilkin07}.  Since the data are not sensitive to the sign of the real part of the scattering length, the quest for a bound state or an unbound pole can not be answered. The pole position or binding energy is
\begin{equation}
|Q_{\rm{^3He}\eta } | \approx 0.30\,\,{\rm{MeV}}.
\end{equation}
From the model calculations it is known that binding is more probable for heavier nuclei than for lighter nuclei. Indeed in a recent study of the $s$ wave in $d\,d\to\eta\,\alpha$ reaction \cite{Budzanowski09b} employing a tensor polarized deuteron beam a scattering length $a_{{}^4He\eta }  = \left[ { \pm \left( {3.1 \pm 0.5} \right) + i \cdot \left( {0 \pm 0.5} \right)} \right]{\rm{fm}}$ was found yielding a pole position \begin{equation}|Q_{^4He\eta }|\approx 4 \text{ MeV.}\end{equation}
Hence one can expect the binding of $\eta$ mesons with A=7 nuclei to be even stronger.

\section{Experiment}\label{sec:Experiment}

In this paper we present results of a measurement of $\eta$ production on $^6$Li:
\begin{equation}\label{reaction}
p+^6\text{Li}\to\eta+^7\text{Be}.
\end{equation}
Such a measurement was performed earlier at SATURNE \cite{Scomparin93} at a proton beam energy of 683 MeV, corresponding to a momentum of 1322 MeV/c or to an excess energy of 19.13 MeV. Two photons were measured with a two-arm spectrometer. In total eight events were detected with three being believed to stem from background. A differential cross section of  $d\sigma/d\Omega= (4.6\pm 3.8)$ nb/sr was reported. This value corresponds to the sum of the ground and all excited states of $^7$Be up to about $\approx$ 10 MeV excitation. These are the particle bound states with $L=1$: $3/2$ (g.s.) and $1/2$ (0.43 MeV), and particle unbound states with $L=3$: $7/2$ (4.57 MeV) and $5/2$ ($\approx$ 7 MeV) \cite{Tilley02}. The reaction was theoretically studied by assuming a reaction $p\,d\to \eta^3$He with an additional  $\alpha$ particle as a spectator \cite{Khalili93}. Including the excited states, the experimental cross section could be reproduced. However, the coincidence in values was assumed to be largely fortuitous in view of the large error bars in both the experiment and prediction.

Here we report on an experiment which was conducted even closer to threshold. In contrast to the previous experiment \cite{Scomparin93} we measured the $^7$Be nucleus. This reduces the number of possible excited states to only one since all other excited states are particle unbound. In the next paragraph we will present the experiment. Then we will discuss the result and finally state our conclusions.

\begin{figure}[h]
\begin{center}
\includegraphics[width=0.80\textwidth]{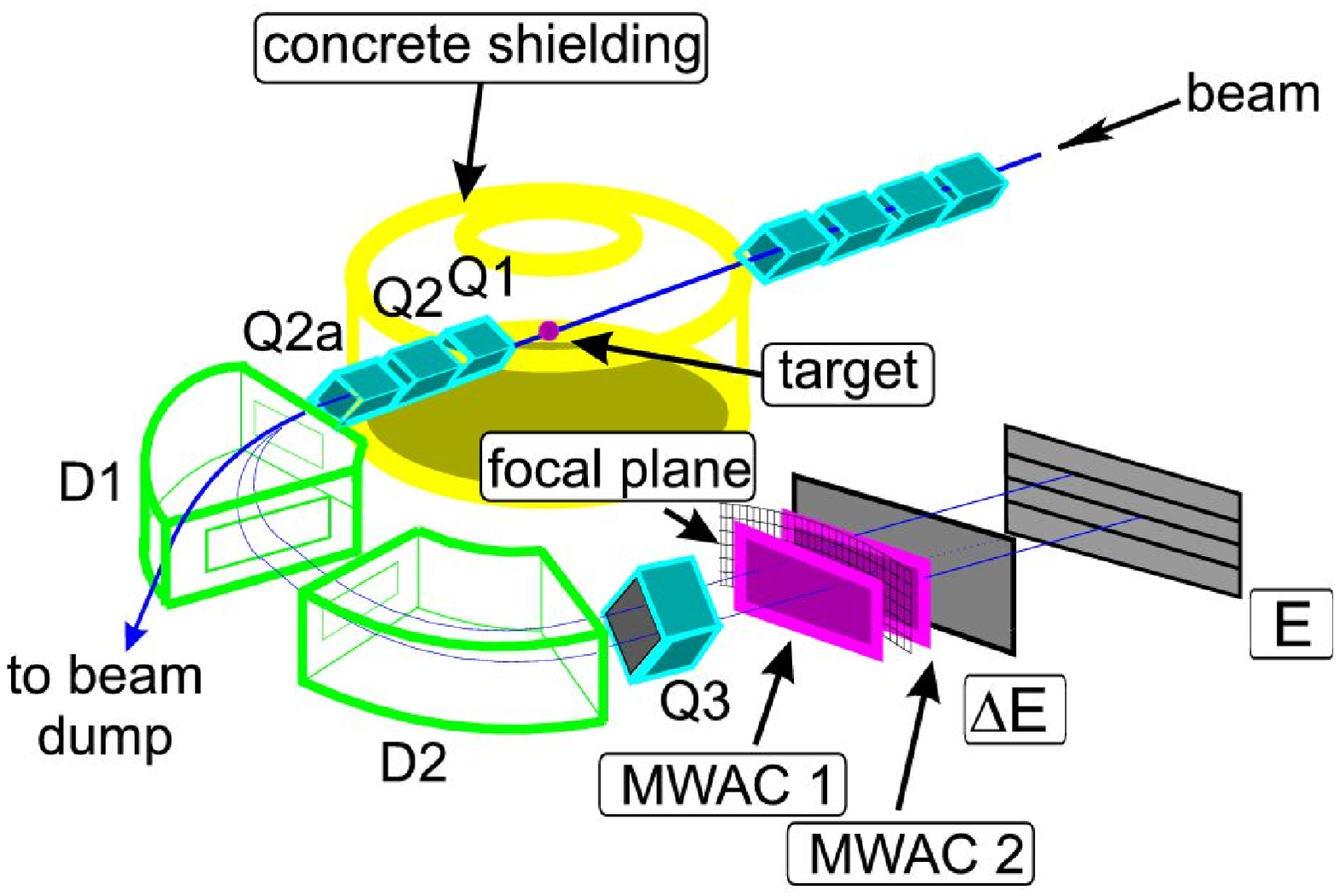}
\caption{(Color online) Schematic view of the Big Karl set up. The beam is focussed with the help of four  quadrupole magnets in the beam line onto the target. With $Q1$, $Q2$, $Q2a$ and  $Q3$ quadrupole magnets in the spectrometer are designated whereas $D1$ and $D2$ denote dipole magnets. MWAC are multi-wire avalanche chambers. $\Delta E$ and $E$ are five scintillators measuring energy and TOF. The focal plane detectors are mounted in a vacuum box which is connected to the vacuum in the magnets.}
\label{Fig:Aufbau}
\end{center}
\end{figure}

The experiment was performed at the cooler synchrotron COSY at J\"{u}lich. A proton beam of 1310 MeV/c momentum corresponding to a beam energy of 673.1 MeV was used. The excess energy for reaction (\ref{reaction}) is 11.28 MeV. The recoiling $^7$Be nuclei were detected with the magnetic spectrograph Big Karl \cite{Drochner98, Bojowald02}. The layout is schematically shown in Fig. \ref{Fig:Aufbau}.
Since the rather low energy Be nuclei are  strongly  ionizing a different set up than the usual one in the focal plane had to be developed. All detectors were placed in a box made of steel.  Vacuum pumps were mounted on the top. It was flanged with its front side to the exit window of the spectrometer. The window has dimensions 65.5$\times$ 6.5 cm$^2$. On both sides there are three flanges which were used to feed the detector signals out of the vacuum. Detectors used to measure the reaction product position and their emission direction were two multi-wire avalanche chambers (MWAC), each of size 70$\times$8 cm$^2$. Each chamber had 546 wires inclined by 45\degr to the left and a similar number inclined to the right. Therefore they were position sensitive in the horizontal as well as vertical direction. Each wire had diameter of 20 $\mu$m, so that only 0.04$\%$ of the chamber area was filled with wire material. The chambers were subdivided into a right and a left half with separate delay line read out. The response of the chambers to ions was tested with beams of $^7$Li at 48 MeV, $^{12}$C at  60 MeV and $^{16}$O at 50 MeV delivered from the BARC-TIFR pelletron accelerator in Mumbai.

The two MWACs were followed by two layers of plastic scintillators. The first one was a bar with dimensions 60$\times$8 cm$^2$. It had a thickness of 0.5 mm and served as $\Delta E$ detector as well as start detector for a TOF measurement. The second layer was a stack of four scintillators of 70$\times$2 cm$^2$, each 2 mm thick. It served as an $E$ detector as well as a stop detector for the TOF measurement. The distance between $\Delta E$ and $E$ detector was 1.02 m, and that between the two MWACs 0.445 m.

The incident beam intensity was measured by calibrated luminosity monitors left and right of the target at large angles. The total number of incident protons was $(6.97\pm 0.70)\times 10^{13}$. The target was a metallic self supporting foil produced by rolling to a thickness of 100 $\mu$m. It was isotopically pure to 99$\%$. Its thickness was a compromise between count rate and energy resolution. It was optimized to 100$\mu$m, corresponding to an energy resolution of 1 MeV. Hence it was not possible to separate and to distinguish between the ground state and the first excited state of $^7$Be.

\section{Results}\label{sec:Results}
The number of $^7$Be nuclei is expected to be small so that particle identification is important to distinguish between frequent background and rare events.
\begin{figure}[!h]
\begin{center}
\includegraphics[width=0.8\textwidth]{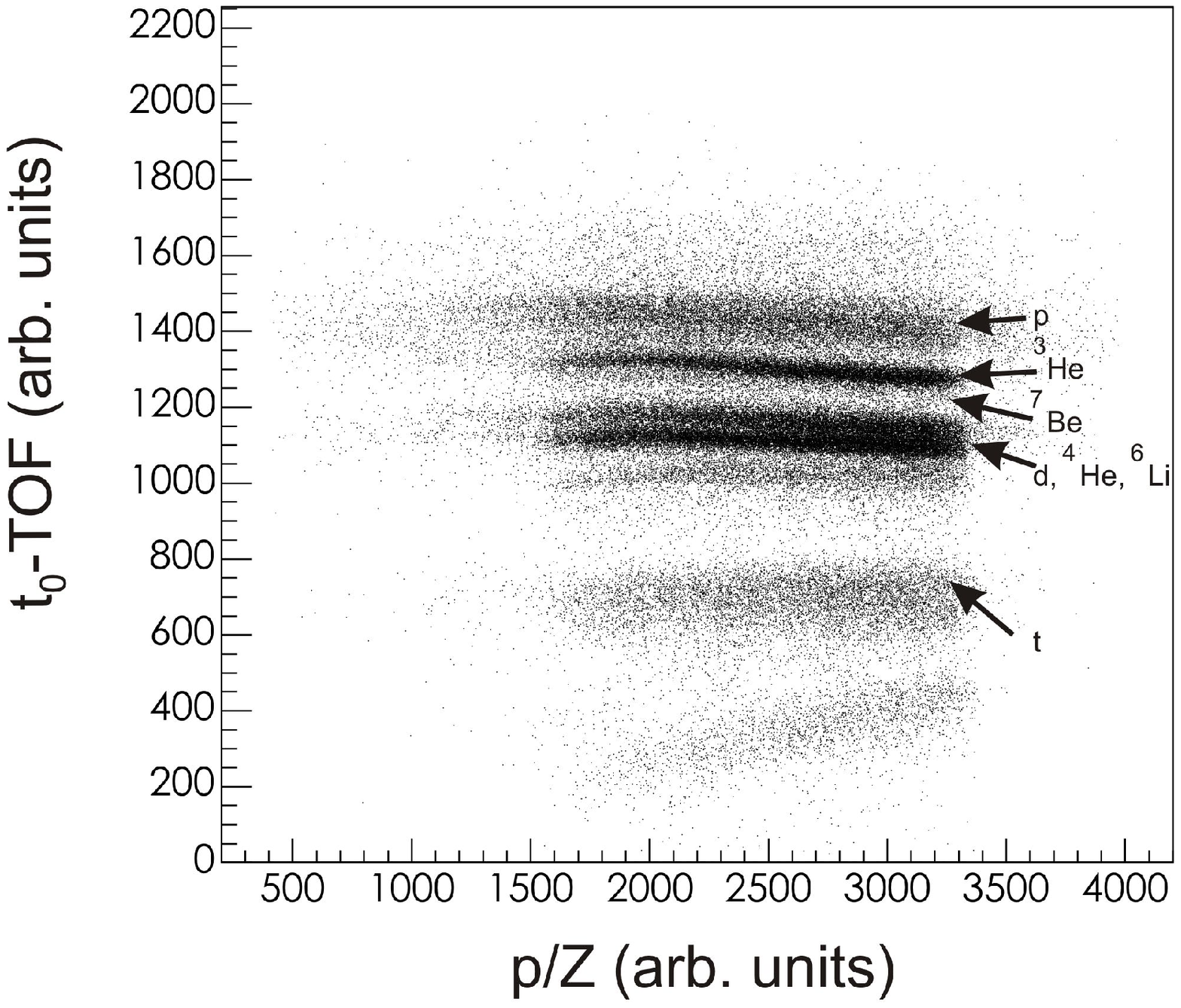}
\caption{The time difference due to time of flight and the momentum of the particles as measured in MWAC~2.}
\label{Fig:TOF}
\end{center}
\end{figure}
Therefore particle identification was performed by redundant methods.

In Fig. \ref{Fig:TOF} the time of flight is shown as a function of the momenta of the particles. The $^7$Be ions were expected to fall  between the bands of deuterons plus $\alpha$-particles and $^3$He.
\begin{figure}[h]
\begin{center}
\includegraphics[width=0.8\textwidth]{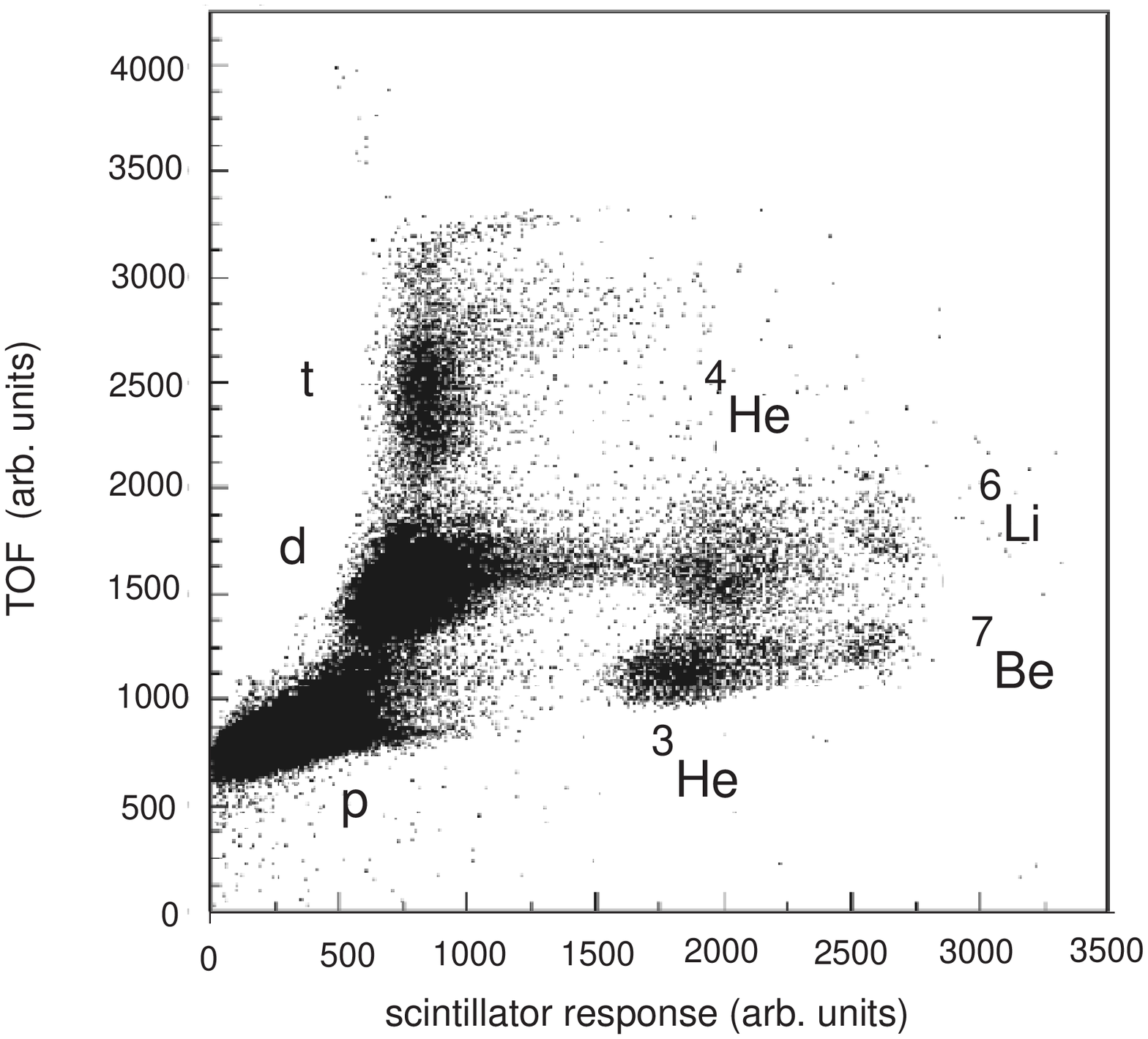}
\caption{Scatter plot of time of flight as function of the output of the $E$ scintillator.}
\label{Fig:TOF-E}
\end{center}
\end{figure}
Fig. \ref{Fig:TOF-E} shows the relation between TOF and energy of the particles. Also on this plot $^7$Be nuclei can be well separated from other species. Similar selections were done for $\Delta E-E$ measurements. The loci for different particles were consistent with Monte Carlo simulations.

\begin{figure}[!h]
\begin{center}
\includegraphics[width=0.8\textwidth]{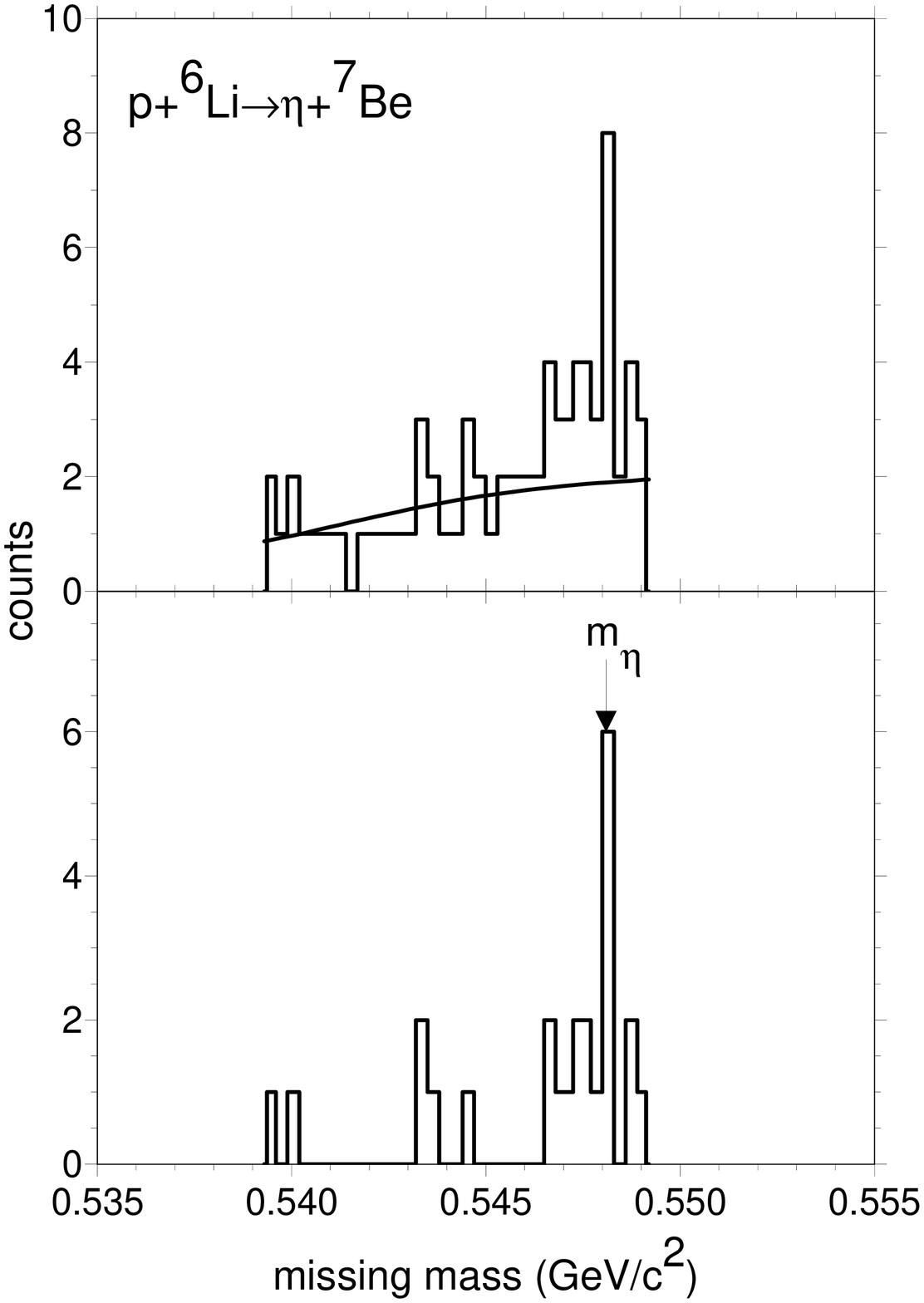}
\caption{Upper panel: Missing mass distribution of the $p+^6$Li$\to ^7$Be+X reaction. The solid curve is the three pion production adjusted to the present data neglecting the peak structure. Lower panel: Missing mass distribution (converted to integers numbers) of the $p+^6$Li$\to \eta + ^7$Be reaction after pion background subtraction.}
\label{Fig:MM}
\end{center}
\end{figure}

\begin{figure}[!h]
\begin{center}
\includegraphics[width=0.8\textwidth]{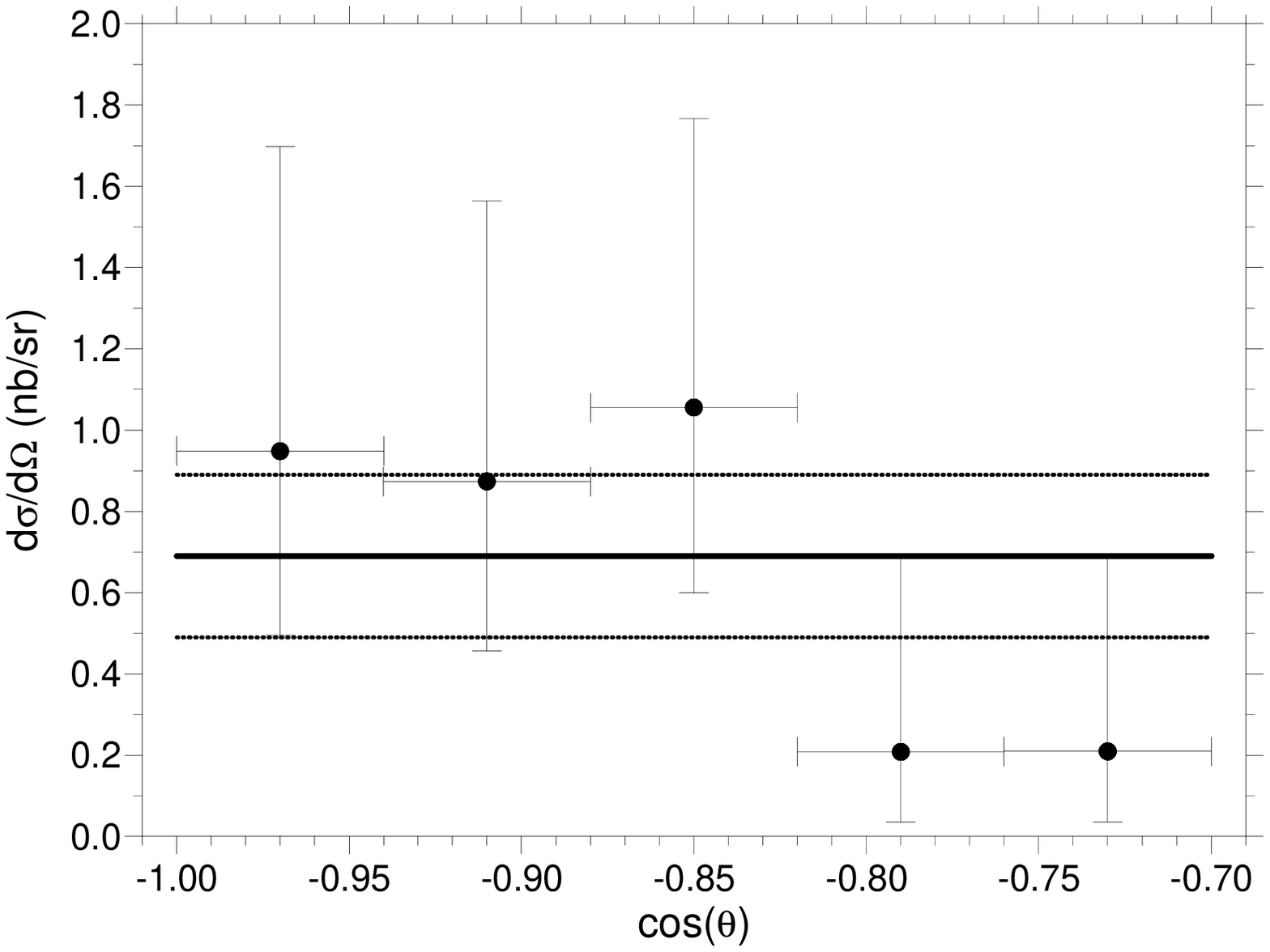}
\caption{Angular distribution of the present measurement. The lines show the mean value together with its uncertainty.}
\label{Fig:ang_dist}
\end{center}
\end{figure}

The obtained missing mass distribution is shown in Fig. \ref{Fig:MM}. Finally simulations were performed for multi-pion production similar to \cite{Budzanowski09b}. The four pion production has no acceptance in the present set-up. Two and three pion production have almost the same missing mass distribution. We then fitted the distribution to the data except to those in the peak area. This physical background was then subtracted and the remaining count rate converted back to integer numbers. This doesn't introduce a large error since the background is almost 2.0 in the peak region. For the events surviving all these cuts and after subtracting background, we obtain the missing-mass distribution which is shown in Fig. \ref{Fig:MM}.
A peak like structure remains containing 15 counts. Fitting the pion background simultaneously with a Gaussian yields 12.7$\pm$5.0 counts. We therefore assume a systematic uncertainty in the total number of counts of 25$\%$. The number of counts can now be converted into cross section. For these events an angular distribution in the c.m. system was deduced which is shown in Fig. \ref{Fig:ang_dist}.
Since we have a small number of counts, we make use of Poisson statistics, yielding asymmetric statistical errors. Systematic uncertainties stem from target thickness (10$\%$), total beam flux (10$\%$), and multi-pion background (25$\%$). All other uncertainties are much smaller. Adding all these uncertainties in quadrature gives for the differential cross section a value
\begin{equation}\label{Result}
\frac{d\sigma}{d\Omega}=(0.69\pm{0.20}\text{( stat.)}\pm 0.20\text{( syst.)})\text{ nb/sr}.
\end{equation}

\section{Discussion}\label{sec:Discussion}
\begin{figure}[h]
\begin{center}
\includegraphics[width=0.8\textwidth]{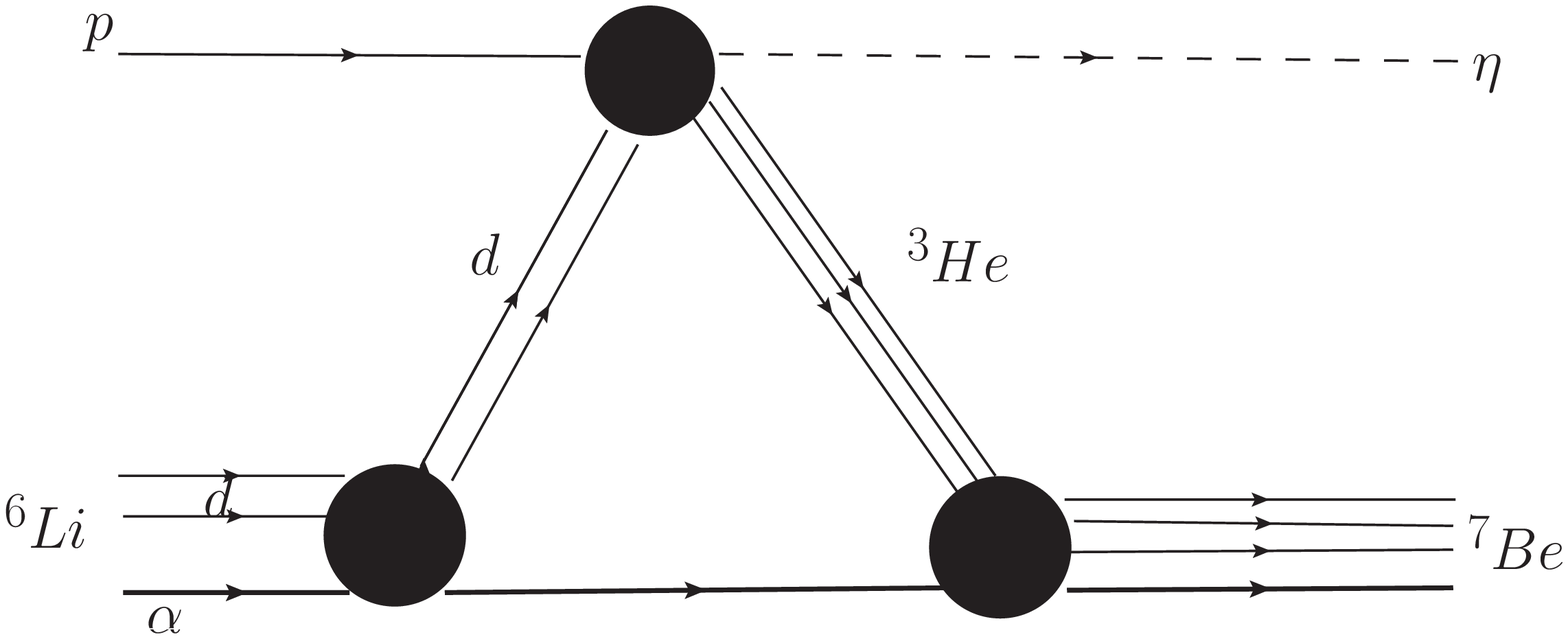}
\caption{$\eta$ production graph treating the $\alpha$ particle (thick line) as a spectator.}
\label{Fig:Direct}
\end{center}
\end{figure}

\begin{figure}[h]
\begin{center}
\includegraphics[width=0.8\textwidth]{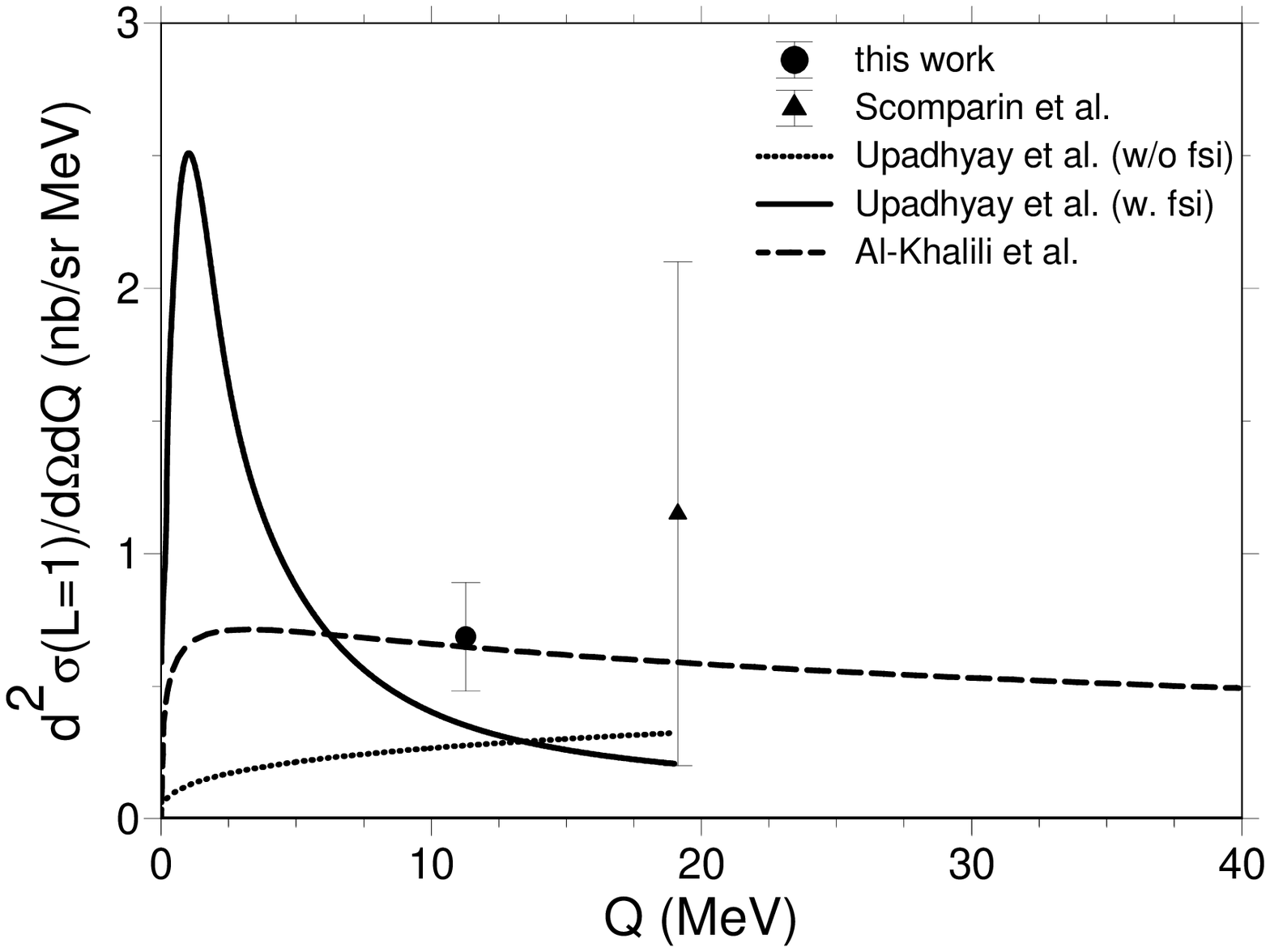}
\caption{Excitation function of the reaction $p+^6\text{Li}\to\eta+^7\text{Be}$ with Be in its ground state and first excited state. The data are the present measurement (full dot) and the for these states corrected result from \cite{Scomparin93} (triangle). The calculations based on the model of \cite{Khalili93} are shown as dashed curve. Those performed by \cite{Upadhyay09} for the total cross section were divided by 4$\pi$. The calculation with a strong \fsi is shown as solid curve while the one without \fsi is shown as dotted curve.}
\label{Fig:exfu}
\end{center}
\end{figure}

This cross section with two possible final states at an excess energy of 11.28 MeV is smaller than the number $4.6\pm 3.8$ nb/sr quoted for an excess energy of 19.13 MeV and four final states \cite{Scomparin93}. In order to make a comparison of the cross sections more meaningful we subtract from the latter experimental result the contributions of the $L=3$ states. Al-Khalili et al. \cite{Khalili93} derived an expression for the differential cross section
\begin{equation}\label{equ:Khalil}
\frac{d\sigma(p^6\text{Li}\to\eta ^7\text{Be})}{d\Omega}= \mathcal{C} \frac{p^*_\eta}{p^*_p}|f(pd\to\eta ^3\text{He})|^2
\sum_j  \frac{2j+1}{2}\mathcal{F}_j^2,
\end{equation}
with $j$ the total angular momentum of the final states in $^7$Be and $\mathcal{F}_j$ their form factors. $\mathcal{C}$ is the overlap of cluster wave functions, $p^*_\eta$ and $p^*_p$ the center of mass momenta of the final and initial system, and $|f|$ the spin averaged matrix element of the underlying more elementary reaction, treating the $\alpha$ particle as a spectator. This reaction is illustrated in Fig. \ref{Fig:Direct}.
We then obtain from Eq. \ref{equ:Khalil}
\begin{equation}\label{equ:L=1}
\frac{d\sigma(L=1)}{d\Omega}=\frac{d\sigma(exp.)}{d\Omega}\frac{\sum_{j=3/2,1/2}\frac{2j+1}{2}\mathcal{F}_j^2} {\sum_{j=3/2,1/2,7/2,5/2}\frac{2j+1}{2}\mathcal{F}_j^2}
\end{equation}
where the experimental cross section contains contributions from L=1 and L=3. The resulting value is compared in Fig. \ref{Fig:exfu} with the present result. The difference between the two experimental results is now much smaller as expected.

Since new data for the underlying $pd\to \eta ^3$He reaction have recently been reported,  we have used all data  in the vicinity of the threshold \cite{Mayer85, Mersmann07, Smyrski07, Adam07} to derive $|f|^2=p^*_p/p^*_\eta \sigma(pd\to\eta ^3\text{He})/d\Omega$ with $p^*_{p,\eta}$ the c.m. momenta of the initial and final state.
\begin{figure}[h]
\begin{center}
\includegraphics[width=0.8\textwidth]{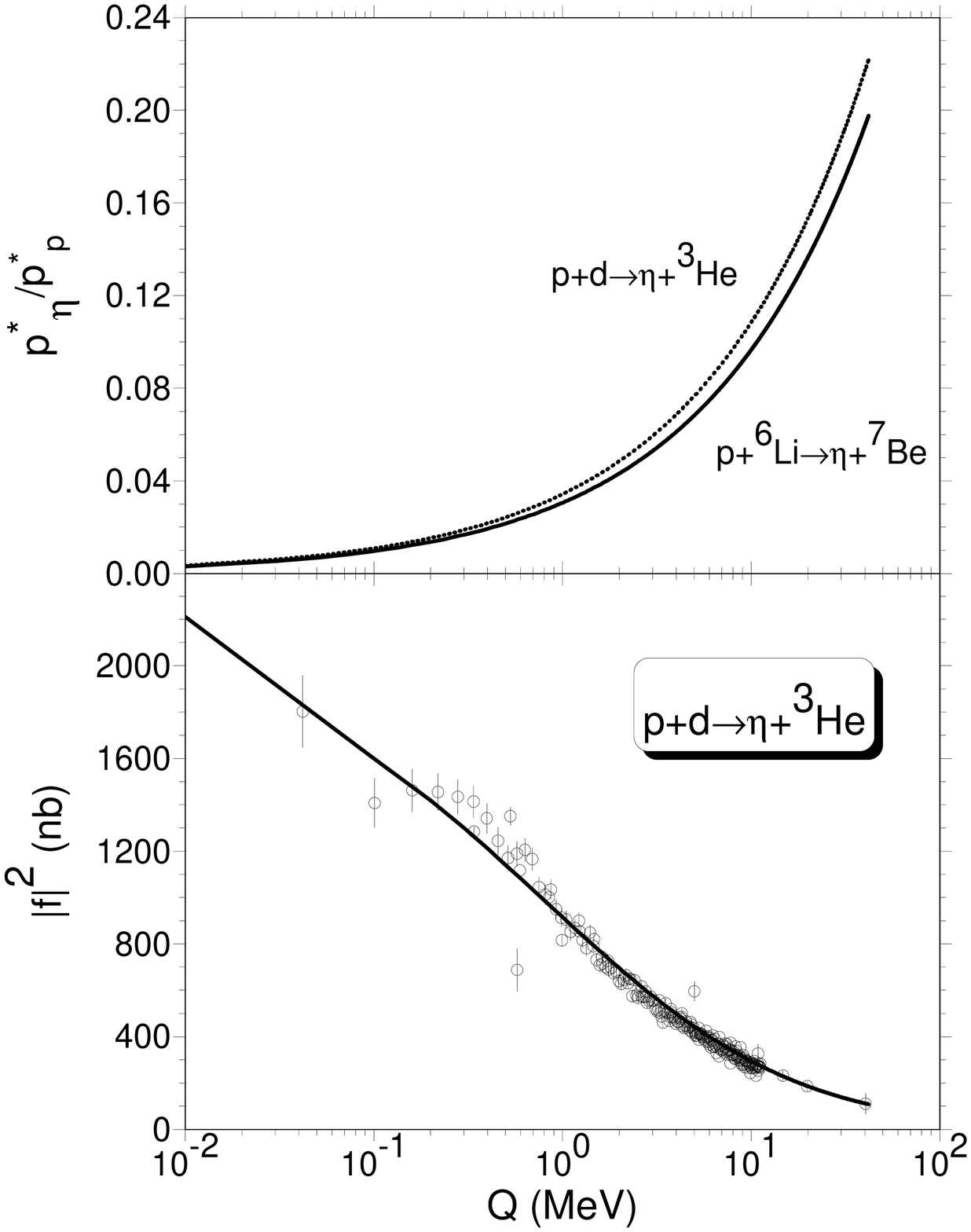}
\caption{The energy dependent quantities entering the model of Al-Khalili et al. \cite{Khalili93}. Upper part: The phase space factor $p^*_\eta/p^*_p$ for the two indicated reactions. Lower part: The spin averaged matrix element $|f|^2=p^*_p/p^*_\eta \sigma(pd\to\eta ^3\text{He})/d\Omega$  as obtained from data \cite{Mayer85, Mersmann07, Smyrski07, Adam07}. The solid curve is a fit to the matrix elements.}
\label{Fig:Matrix}
\end{center}
\end{figure}

The matrix element $|f|^2$ at 19.13 MeV  is now significantly smaller than the value assumed in \cite{Khalili93}. Since the interval of transferred momentum is narrow we ignore its dependence on the overlap integral and on the the form factor and calculate the cross section for the present reaction according to Eq. (\ref{equ:Khalil}).
It is also shown in Fig. \ref{Fig:exfu} as dashed curve. It accounts for the present measurement and meets the error bar of the earlier measurement. It should be mentioned that the calculation is expected to be correct within 66$\%$ due to an uncertainty in $\mathcal{F}_{3/2}^2$.  The shape of the resulting curve is almost the same as the one for the $pd\to\eta^3$He reaction. This is so since the phase factors $p^*_p/p^*_\eta$ for the underlying reaction and the present reaction are almost identical (see Fig. \ref{Fig:Matrix}). This model can therefore not give a decisive answer whether a possible pole in \fsi moves to a larger $Q$-value for the present reaction.

A more recent calculation was reported by the Mumbai group \cite{Upadhyay09}. Again the target and the residual nuclei were treated as being composed of two clusters.
The input is the $\eta$-nucleon interaction where they have assumed a large scattering length in agreement with a bound state. In addition to the graph shown in Fig. \ref{Fig:Direct} they included a rescattering term.  The shape of the excitation function thus becomes different than the one for the underlying $pd\to\eta^3$He reaction.  Their results where the cluster wave functions were generated by Woods-Saxon potential are also shown in Fig. \ref{Fig:exfu}. Here we have divided their result for the total cross section by $4\pi$. This calculation with a rather large $\eta$-nucleon scattering length differs largely from the one within the model of \cite{Khalili93}. The calculation without \fsi, which is phase space behavior, shows the energy dependence of the data but underestimates the measured data. A small \fsi can not be ruled out.

In summary we have measured the momentum $\vec{p}$ of $^7$Be nuclei from the reaction $p+^6\text{Li}\to\eta+^7\text{Be}$ at a beam momentum of 1310 MeV/c with the high resolution magnetic spectrograph Big Karl. Dedicated focal plane detectors were developed and used: MWACs and $\Delta E-E$ scintillators. The latter permitted time-of-flight measurements. All detectors were working in vacuum. A differential cross section of 0.69 nb/sr in the c.m. system was obtained which corresponds to a total cross section of $(8.6\pm 2.6\text{ stat.}\pm 2.4\text{ syst.})$ nb when isotropic emission is assumed. This cross section  at an excess energy of 11.28 MeV is almost an order of magnitude smaller than the number $4.6\pm 3.8$ nb/sr quoted for an excess energy of 19.13 MeV \cite{Scomparin93}.  However, in the present experiment only two possible final states exist corresponding to angular momentum states with only $L=1$ while in the earlier experiment four final states with $L=1$ and $L=3$ contribute. Comparison for only $L=1$ states reduces the difference. The data were compared with model calculations. Although the calculations predict the right order of magnitude one can not distinguish the size of the \fsi. More data especially closer to threshold are necessary to pin down the open problems.

The reactions of $\eta$ production on light nuclei with  two-body final states as discussed in section \ref{sec:Introduction} and in this work have been performed at energies below the $\eta$ production threshold of proton-proton interactions.  Different scenarios have been applied to account for such processes like multi-step processes, interaction of the projectile with a nucleon having large Fermi momentum and coherent interaction. Thus subthreshold $\eta$ production on light nuclei is interesting in itself. In addition $\eta$ production differs from $\pi^0$ production due to its coupling to a resonance (the $N^*(1535)$). We compare the present total cross section with those for other light nuclei reactions for about the same excess energy. The values are 407$\pm$20 nb for $p+d\to\eta +^3$He reaction\cite{Mersmann07}, 16$\pm$1.6 nb for the $d+d\to\eta+^4$He reaction \cite{Budzanowski09b}, and for the present $p+^6\text{Li}\to\eta+^7\text{Be}$ reaction: $(8.6\pm 2.6\text{ stat.}\pm 2.4\text{ syst.}$ nb. This shows a dramatic decrease of the cross section with increasing mass number for the proton induced reactions. The center of mass momenta in all three reactions are compatible. This is in strong contrast to inclusive production in heavy ion collisions where a dependence $\sigma\propto (A_pA_t)^{2/3}$ was found \cite{Noll}. So the origin of the decrease will be the A-dependence of the corresponding form factors at the large momentum transfer of 800 to 900 MeV/c. This reflects the fact that it is more unlikely to fuse to a heavy system than to a light system.

\begin{acknowledgments}
We are grateful to the COSY crew for preparing an excellent beam. One of us (H. M.) acknowledges C. Wilkin for fruitful discussions.  We appreciate the support received from the European community research
infrastructure activity under the FP6 ``Structuring the European
Research Area'' programme, contract no.\ RII3-CT-2004-506078, from
the Indo-German bilateral agreement, from the Bundesministerium f\"{u}r Bildung und Forschung, BMBF (06BN108I), from the Research Centre
J\"{u}lich (FFE), and from GAS Slovakia (1/4010/07).
\end{acknowledgments}



\end{document}